\documentclass[aps,pra,twocolumn,superscriptaddress]{revtex4-2}

\usepackage{graphicx}
\usepackage{multirow}
\usepackage{amsmath,amssymb,amsfonts}
\usepackage{amsthm}
\usepackage{mathrsfs}
\usepackage{xcolor}
\usepackage{textcomp}
\usepackage{booktabs}
\usepackage{algorithm}
\usepackage{algorithmicx}
\usepackage{algpseudocode}
\usepackage{listings}
\usepackage{siunitx}
\usepackage{dirtytalk}
\usepackage{xr}
\usepackage{braket}
\usepackage{placeins}
\usepackage[hidelinks]{hyperref}
\usepackage[separate-uncertainty = true, multi-part-units=single]{siunitx}
\usepackage{xcolor}

\begin{document}

\title{Optimising germanium hole spin qubits with a room-temperature magnet}

\author{Cécile X. Yu}
\thanks{These authors contributed equally to this work.}
\affiliation{QuTech and Kavli Institute of Nanoscience, Delft University of Technology, P.O. Box 5046, 2600 GA Delft, The Netherlands}

\author{Barnaby van Straaten}
\thanks{These authors contributed equally to this work.}
\affiliation{QuTech and Kavli Institute of Nanoscience, Delft University of Technology, P.O. Box 5046, 2600 GA Delft, The Netherlands}

\author{Alexander S. Ivlev}
\affiliation{QuTech and Kavli Institute of Nanoscience, Delft University of Technology, P.O. Box 5046, 2600 GA Delft, The Netherlands}

\author{Valentin John}
\affiliation{QuTech and Kavli Institute of Nanoscience, Delft University of Technology, P.O. Box 5046, 2600 GA Delft, The Netherlands}

\author{Stefan D. Oosterhout}
\affiliation{QuTech and Netherlands Organisation for Applied Scientific Research, 2628 CK Delft, The Netherlands}

\author{Lucas E. A. Stehouwer}
\affiliation{QuTech and Kavli Institute of Nanoscience, Delft University of Technology, P.O. Box 5046, 2600 GA Delft, The Netherlands}

\author{Francesco Borsoi}
\affiliation{QuTech and Kavli Institute of Nanoscience, Delft University of Technology, P.O. Box 5046, 2600 GA Delft, The Netherlands}

\author{Giordano Scappucci}
\affiliation{QuTech and Kavli Institute of Nanoscience, Delft University of Technology, P.O. Box 5046, 2600 GA Delft, The Netherlands}

\author{Menno Veldhorst}
\email{M.Veldhorst@tudelft.nl}
\affiliation{QuTech and Kavli Institute of Nanoscience, Delft University of Technology, P.O. Box 5046, 2600 GA Delft, The Netherlands}

\date{\today}

\begin{abstract} 

    Germanium spin qubits exhibit strong spin-orbit interaction, which allow for high-fidelity qubit control, but also provide a strong dependence on the magnetic field. Superconducting vector magnets are often used to minimize dephasing due to hyperfine interactions and to maximize spin control, but these compromise the sample space and thus challenge scalability. Here, we explore whether a permanent magnet outside the cryostat can be used as an alternative. Operating in a hybrid mode with an internal and external magnet, we find that we can fine-tune the magnetic field to an in-plane orientation. We obtain a qubit dephasing time $T_2 ^{*}=$ \SI{13}{\us}, Hahn-echo times $T_2^\mathrm{H} =$ \SI{88}{\us}, and an average single-qubit Clifford gate fidelity above 99.9\%, from which we conclude that room temperature magnets allow for high qubit performance. Furthermore, we probe the qubit resonance frequency using only the external magnet, with the internal superconducting magnet switched off. Our approach may be used to scale semiconductor qubits and use the increased sample space for the integration of cryogenic control circuitry and wiring to advance to large-scale quantum processors.
    
 \end{abstract}

\maketitle

\section{Introduction}

Spin qubits in semiconductor quantum dots are an attractive platform for quantum computing, as they offer long coherence times, allow for high-fidelity quantum gates, and have high compatibility with industrial semiconductor fabrication \cite{Burkard2023SemiconductorQubits}. Germanium has emerged as a promising material for hosting hole-based spin qubits, offering strong spin-orbit interaction and all-electrical control \cite{Watzinger2018AQubit,Hendrickx2021AProcessor, Wang2024OperatingSpins, Scappucci2020TheRoute}. Despite these advantages, achieving precise control over the qubit properties requires accurate alignment of the external magnetic field to the substrate plane to optimise coherence and fidelity due to the strong $g$-tensor anisotropy \cite{Hendrickx2024Sweet-spotSensitivity, Mauro2024GeometryQubits, Saez-Mollejo2025ExchangeQubits, John2024AControl}.

Even if the $g$-tensor plane is aligned with the sample crystallographic axes, achieving sub-degree accuracy in magnetic field alignment is challenging due to factors such as sample mounting variability and the differential thermal contraction of materials within the cryostat. While vector magnets can compensate for these misalignments to minimise hyperfine-induced dephasing and enhance the efficiency of electric dipole spin resonance (EDSR) operations \cite{Hendrickx2024Sweet-spotSensitivity, Bassi2024OptimalQubits}, their implementation requires considerable space and thermal load within the cryostat, and limit the available sample space \cite{DaPrato2025Step-by-stepMagnet}. 

\begin{figure}[ht!]
	\includegraphics{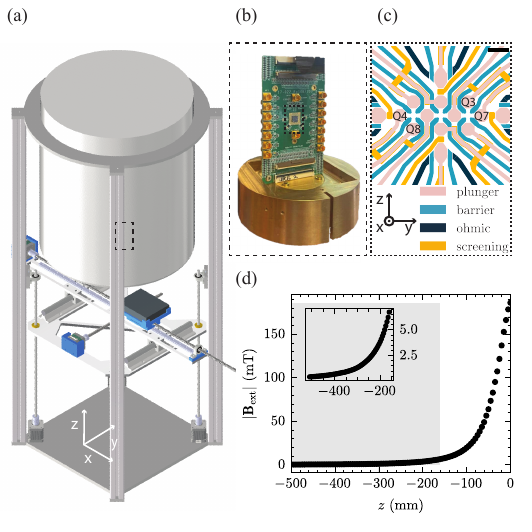}
	\caption{\textbf{Experimental set-up.} \textbf{a}, Diagram showing the XYZ stage used to position the permanent magnet (grey block). The stage is attached to the vacuum can of the fridge. 
    \textbf{b,} Picture of the qubit chip mounted on the printed circuit board inside the cryostat which defines the origin of the coordinate system. The copper piece below the PCB corresponds to the bore of the uniaxial superconducting magnet limiting the sample space. 
    \textbf{c,} Device layout indicating the qubits investigated in the experiments. The scale bar corresponds to \SI{200}{nm}.
    \textbf{d,} Magnetic field amplitude of the permanent magnet as a function of distance. The grey shaded area denotes the $z$-range used in the experiments and the inner plot shows the field strength in this $z$-range.
    }
	\label{fig: fridge}
\end{figure}

This work explores an alternative approach using a movable permanent magnet outside the dilution refrigerator to fine-tune the magnetic field at the qubit location in addition to the uniaxial superconducting magnet in the cryostat. This internal magnet provides a primary magnetic field of a few tens of millitesla, chosen to be small enough that any misalignment can be effectively corrected using the external magnet.

To characterise the feasibility of this approach, we perform EDSR drive on the qubit to evaluate its resonance frequency as a function of the uniaxial superconducting magnetic field strength, as well as the position of the external magnet relative to the sample. This study is further extended by characterising the qubit performance through Ramsey and Hahn echo experiments, along with single-qubit gate randomized benchmarking and gate set tomography. Our findings demonstrate that a qubit fine-tuned by an external magnet can achieve high quality qubit control, with fidelities exceeding 99.9\%.
In this work, a uniaxial superconducting magnet inside the fridge supplies the primary millitesla magnetic field. Nevertheless, we also demonstrate qubit control when the superconducting magnet is turned off, giving the possibility to entirely replace the superconducting magnet by a small permanent one close to the sample, offering promising applications for large-scale quantum processor architectures. 

\section{Experimental set-up}

In this experiment, we use a standard cryogen-free dilution refrigerator equipped with a uniaxial superconducting magnet. To fine-tune the magnetic field, we move an NdFeB N45 block magnet in three dimensions beneath the fridge (see \autoref{fig: fridge}a). The motion of the external magnet is accomplished by suspending a Cartesian gantry system from the vacuum can of the dilution refrigerator, with motion along the $x$, $y$, and $z$ axes controlled remotely by linear actuators and an Arduino board. 

\begin{figure}[ht!]
	\includegraphics{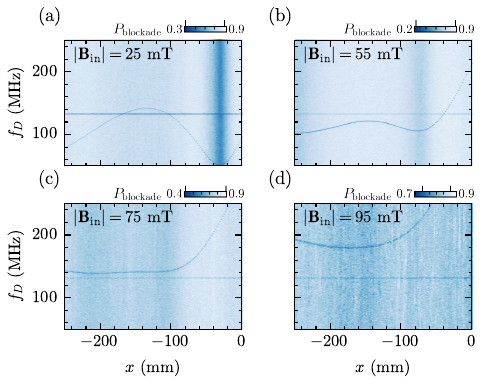}
	\caption{\textbf{Effect of the internal magnetic field magnitude.} Microwave spectroscopy of the qubit frequency as a function of permanent magnet position in \textit{x} at different internal magnetic field magnitude. $P_\mathrm{blockade}$ defines the probability of the even spin state being blockaded and $f_\mathrm{D}$ is the microwave drive tone frequency. The resonance around \SI{130}{MHz} corresponds to the resonance frequency of the readout resonator.}
    \label{fig:field}
\end{figure}

For our coordinate system, we take the $z$ axis as parallel to the vertical axis of the dilution refrigerator. If the sample were mounted perfectly, fields applied in this direction correspond to the $[110]$ crystallographic direction and therefore are in-plane. In previous work on this device \cite{AStehouwer2025ExploitingMicron-scale}, we estimate there is a 2-3 degree misalignment, meaning that the magnetic field of the internal superconducting solenoid has a small out-of-plane component. We define the $x$ axis close to parallel to the heterostructures $[001]$ growth direction; therefore, fields applied in this direction are out of plane. By moving the magnet in the $x$ direction, we expect to be able to compensate and cancel out the orthogonal field component, provided the external magnetic field is sufficiently strong. The coordinate system is defined relative to the sample, meaning that $(x, y, z) = (0, 0, 0)$ corresponds to the position of the sample, see \autoref{fig: fridge}b and c. 

We perform our qubit experiments on a 10-quantum dot device on a Ge/SiGe heterostructure as detailed in Ref. \cite{AStehouwer2025ExploitingMicron-scale,John2024AControl} and investigate the effect of the magnetic field. We focus on qubits Q8 and Q3, which are initialised and read out using ancilla qubits Q4 and Q7, respectively.
We will refer to the magnetic field applied by the superconducting uniaxial solenoid inside the fridge as $\mathbf{B}_\mathrm{in}$, and $\mathbf{B}_\mathrm{ext}$ as the magnetic field generated by the external magnet. In \autoref{fig: fridge}d, we probe $|\mathbf{B_\mathrm{ext}}|$ as a function of distance $z$ using 3D magnetic Hall sensors. The closest $z$-position used in the experiments is \SI{-160}{mm}, which corresponds to an external field strength of $\approx$ \SI{6.2}{mT}, assuming no screening effects from the uniaxial magnet.

\section{Qubit frequency tunability}

\begin{figure}[ht!]
	\includegraphics{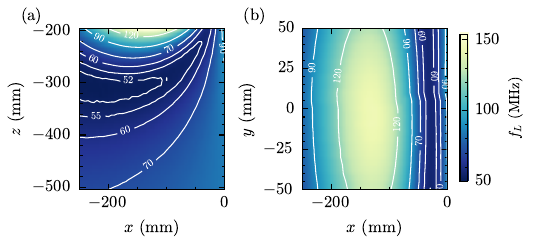}
	\center
	\caption{\textbf{Experimentally measured Larmor frequencies of the qubit as a function of magnet position at $|\mathbf{B_\mathrm{in}}| = 25$ mT.} \textbf{a,} Larmor frequency in the \textit{xz}-plane of the permanent magnet at $y =$ \SI{0}{mm}. \textbf{b,} Larmor frequency in the \textit{xy}-plane of the permanent magnet at $z =$ \SI{-200}{mm}. The contour lines are defined by connecting the corresponding experimental data points.}
    \label{fig:larmor}
\end{figure}

\begin{figure*}
    \centering
	\includegraphics{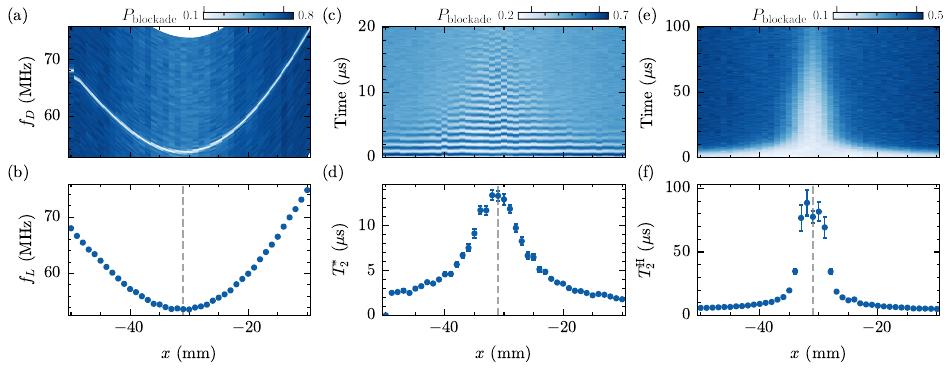}
	\caption{\textbf{Qubit coherence times at the magnetic field sweet-spot.} \textbf{a, b,} Larmor frequency $f_\mathrm{L}$ as a function of the \textit{x}-position of the permanent magnet extracted from the microwave spectroscopy. \textbf{c, d,} Dephasing times $T_2^*$ as a function of the magnet position in $x$ measured with a Ramsey sequence. \textbf{e, f,} $T_2^\mathrm{H}$ measured with a Hahn echo sequence. The dashed grey lines indicate the optimal field spot at the minimum of the Larmor frequency where the magnetic field points in-plane. The error bars represent the precision of the fit.}
    \label{fig:coherence}
\end{figure*}

We begin by assessing the effect of the external magnet on the total magnetic field sensed by the spin qubit at different internal magnetic field amplitudes applied on the uniaxial superconducting magnet. To do so, we perform EDSR drive on the qubit Q8 and map its Larmor frequency as a function of the magnet position in $x$ for different $|\mathbf{B}_\mathrm{in}|$ as shown in \autoref{fig:field}. The external magnet is placed at $(y,z) = (0, -200)$ mm and the \textit{x}-position varies from 0 to \SI{-250}{mm}. As the external magnet is displaced along the \textit{x}-axis, both the effective $g$-factor and the orientation of the total magnetic field experienced by the qubit change, resulting in variations in the Larmor frequency.

At low internal magnetic fields, we observe that the qubit resonance frequency has both a minimum and a local maximum as a function of the magnet position, see \autoref{fig:field}a and b. We can attribute the minimum to the point where the magnetic field generated by the external magnet cancels the out-of-plane component of the solenoid. Due to the large $g$-tensor anisotropy \cite{Scappucci2020TheRoute}, this translates into the magnetic field being aligned with the principal in-plane direction of the hole $g$-tensor, where the qubit $g$-factor and Larmor frequency are at a minimum. Continuing to move further off-axis, the external magnetic field becomes increasingly out-of-plane and begins to overcompensate the internal field. However, at the same time, the qubit experiences a smaller magnetic field of the permanent magnet due to it moving further away from the sample. The competition of these two effects results in a qubit Larmor frequency that increases to a local maximum before beginning to decrease again. At larger internal magnetic fields (see \autoref{fig:field}c and d), the qubit resonance frequency exhibits a parabolic dependence on the position of the magnet. In this regime, the external field is not strong enough to fully compensate for the out-of-plane component of the internal field. Note that in \autoref{fig:field}a and b, we observe that the qubit initialisation and readout fidelity degrades from out-of-plane to in-plane magnetic fields. We attribute this degraded initialisation of the spins in the $\ket{\uparrow\downarrow}$ state, due to in-plane magnetic fields increasing the quantization axis difference between the qubits, enhancing the probability to prepare the $\ket{T_-} = \ket{\downarrow\downarrow}$ \cite{Jirovec2022DynamicsDifferences}.

Now we set $|\textbf{B}_\mathrm{in}| = $ \SI{25}{\milli\tesla} and fully characterise the tunability of the Larmor frequency in the $xz$- and $xy$-planes as shown in \autoref{fig:larmor}. The Larmor frequency, $f_\mathrm{L}$, varies from 50 MHz to 150 MHz, with the \textit{x}-axis playing the dominant role by determining the out-of-plane component of the total magnetic field, and thus causing the strongest tunability on the qubit frequency. In the \textit{xz}-plane, the qubit resonance frequency exhibits a minimum corresponding to the optimal field orientation where the external magnet compensates the internal magnetic field, resulting in a fully in-plane configuration. In contrast, in the \textit{xy}-plane, variations in magnet position primarily tune the in-plane magnetic field orientation.

\section{Qubit coherence times}

We now focus on the qubit performance at the magnetic field sweet spot to enhance the qubit dephasing times. We fix the \textit{z}-position to \SI{-200}{mm} and the \textit{y}-position to zero, while the \textit{x}-position is varied around the point corresponding to the minimum Larmor frequency. At this position, the magnetic field is expected to be aligned within the plane of the sample, where dephasing noise is minimized due to the hyperfine interactions with nuclear spins being minimal \cite{Fischer2008SpinDot, Hendrickx2024Sweet-spotSensitivity}. 
Thus, to enhance the qubit dephasing times, the magnetic field has to be aligned perpendicular to the growth direction, where the anisotropic hyperfine interaction between heavy hole states and nuclear spins is expected to be negligible \cite{Fischer2008SpinDot}.

We measure the coherence times of Q8 by means of Ramsey and Hahn echo experiments to probe the impact of field tuning on spin coherence as shown in \autoref{fig:coherence}. 
We find that when the Larmor frequency reaches its minimum value, $T_2^*$ and $T_2^\mathrm{H}$ are simultaneously maximised, indicating an in-plane magnetic field orientation.
By orienting the magnetic field fully in-plane at the hyperfine noise sweet spot, the dephasing time of the qubit reaches $T_2^*$ = \SI{13.41(53)}{\us} while the echo time $T_2^\mathrm{H}$ = \SI{88.77(9.99)}{\us}. This is comparable with coherence times measured at the magnetic field sweet spot in a vector magnet \cite{Hendrickx2024Sweet-spotSensitivity}. 
Qubit Q3 also exhibits the same trend as a function of the external magnet position along the $x$ direction, see Supplementary Note 1. 
For comparison, without the external magnet, the Ramsey and Hahn echo experiments on Q8 (see Supplementary Note 2) indicate coherence times of $T_2^*$ = \SI{1.70(12)}{\us} and $T_2^\mathrm{H}$ = \SI{4.23(11)}{\us}, respectively, revealing a hyperfine noise limited regime \cite{AStehouwer2025ExploitingMicron-scale}.  
In Supplementary Note 3, we also show the Rabi drive efficiency of Q8 as a function of the magnet $x$-position. It reveals a peak in drive efficiency that coincides with the maximum coherence time, indicating that, for this qubit, an in-plane magnetic field is beneficial for both long coherence times and fast driving \cite{Hendrickx2024Sweet-spotSensitivity, Mauro2024GeometryQubits}.

We furthermore evaluate the single-qubit gate fidelity using both randomized benchmarking (RB) and gate set tomography (GST). We perform RB using the Clifford set detailed in Supplementary Note 4, giving us an average Clifford fidelity of $F_\mathrm{C} =$ \SI{99.936(5)}{\%}. By dividing the average gate fidelity by the number of native gate per Clifford operation, we obtain a single-qubit gate fidelity of \SI{99.980(2)}{\%}, comparable to the highest reported values for qubits and measured using vector magnets \cite{Lawrie2023SimultaneousThreshold, Hendrickx2024Sweet-spotSensitivity}. 
GST, which is more sensitive to systematic errors, reports average fidelities of $99.95 \pm 0.02$\% for the $X90$ gate and $99.88 \pm 0.02$\% for the $Y90$ gate (see Supplementary Note 5). The lower fidelities reported by GST are expected to come from dynamical decoupling effects present in the RB sequence \cite{Dehollain2016OptimizationTomography}.

\section{Zero internal field operation}
A key challenge in scaling towards large-scale quantum processors is represented by the limited space available at the base temperature space in a cryostat \cite{Mohseni2024HowQubits}. By removing the superconducting magnet inside the cryostat, the available sample space can be increased substantially, allowing for more control lines and improved prospects of scalability. 

\begin{figure}
	\includegraphics{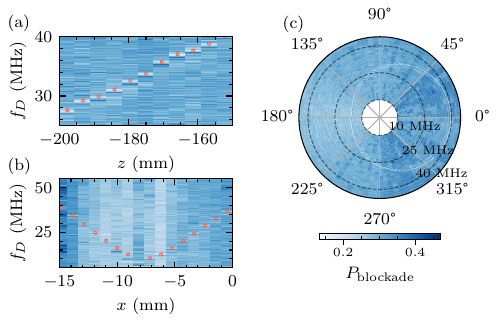}
	\caption{\textbf{Zero internal field Larmor frequency dependence.} \textbf{a,} Microwave spectroscopy of the qubit frequency as a function of $z$-position of the external magnet while keeping $(x,y) =(0,0)$.  
    \textbf{b,} Microwave spectroscopy of the qubit frequency as a function of $x$-position of the external magnet showing a minium value where the field is most likely in plane. $y$ is kept at zero and $z =$ \SI{-160}{mm}. The discrepancy between the Larmor frequency at $x=$ \SI{0}{mm} in plots \textbf{a} and \textbf{b} may be caused  by imprecisions in the magnet movement (more details in Supplementary Note 6). The red circles highlight the maximum value at each position. \textbf{c,} Microwave spectroscopy as a function of circular movement of the external magnet in the $xy$-plane at $z=$ \SI{-200}{mm}. The measurement at \SI{0}{\degree} corresponds to $x=$ \SI{5}{mm} and $y=$ \SI{0}{mm}.
    }
    \label{fig:zero_field}
\end{figure}

Here we simulate this scenario by switching off the internal magnetic field, such that the qubit only senses the magnetic field of the permanent magnet. In this configuration, we are able to probe the Zeeman splitting energy of the qubit Q3 as a function of the external magnet position see \autoref{fig:zero_field}. 
By moving the magnet along the $x$ direction, the qubit Larmor frequency varies from \SI{40}{MHz} to \SI{10}{MHz} due to the strong anisotropy of the $g$-tensor and the variation in the magnetic field magnitude. As the quantization axis can significantly change for magnetic fields close to in-plane, this strongly affects the readout and initialisation visibility. Nevertheless, this suggests the possibility of operating spin qubits without the need for a superconducting magnet inside the cryostat. Bringing the sample closer to the bottom of the fridge and by completely removing the superconducting magnet, which may screen the external magnetic field, we predict that magnetic fields of several tens of millitesla at the sample can be created. 

To achieve even higher magnetic fields at the device level up to \SI{1}{T}, a block magnet can be placed inside the cryostat near the qubit chip \cite{Adambukulam2021AnTemperatures,Rooney2025GateGermanium}. This can be combined with a room-temperature permanent magnet to fine-tune the magnetic field orientation for optimal qubit operations. We note that in natural germanium, the longest coherence is observed when the magnetic field direction is pointing in-plane \cite{Hendrickx2024Sweet-spotSensitivity}. However, the optimal magnetic field direction may point away from the perfect in-plane direction for isotopically purified germanium, as hyperfine interaction is reduced. The magnetic field can therefore be directed a few degrees out-of-plane to optimize other factors, such as the charge-noise sensitivity, qubit control, and uniformity \cite{John2024AControl}. 

\section{Conclusion}

Superconducting magnets inside dilution refrigerators require considerable space, and can introduce current noise or ground loops into the qubit system.
We demonstrate that a room-temperature permanent magnet offers an effective alternative to fine-tune the magnetic field experienced by germanium spin qubits. 
Our versatile room-temperature magnet setup enables in-plane magnetic field alignment, allowing natural germanium devices to operate in regimes less affected by hyperfine noise, thereby enhancing qubit coherence and gate fidelity.
Even qubits hosted in isotopically purified materials \cite{Moutanabbir2024NuclearWells, Chekhovich2013NuclearDots} will need to be tuned to optimal operating conditions \cite{John2024AControl}, which can similarly done with the external magnet approach presented here. The cryostat space may then be used to integrate control wiring and cryo electronics for the construction of large-scale quantum hardware.

\section*{Acknowledgements}

We are grateful to Jason Mensingh for providing the external magnet, Nico Alberts and Tim Hiep for the fine precision mechanics.
This research was supported by the European Union through the Horizon 2020 research and innovation programme under the Grant Agreement No. 951852 (QLSI) and the Horizon Europe Framework Programme under grant agreement No. 101069515 (IGNITE). F.B. and M.V. acknowledges support from the NWO through the National Growth Fund program Quantum Delta NL (grant NGF.1582.22.001). This research was sponsored in part by the Army Research Office (ARO) under Award No. W911NF-23-1-0110 and by The Netherlands Ministry of Defence under Awards No.QuBits R23/009. The views, conclusions, and recommendations contained in this document are those of the authors and are not necessarily endorsed nor should they be interpreted as representing the official policies, either expressed or implied, of the Army Research Office (ARO) or the U.S. Government, or The Netherlands Ministry of Defence. The U.S. Government and The Netherlands Ministry of Defence are authorized to reproduce and distribute reprints for Government purposes notwithstanding any copyright notation herein.

\section{Data availability}
The data and analysis code supporting the findings of this study are openly available in the 4TU.ResearchData repository under \url{https://doi.org/10.4121/eb6053c1-e70c-401c-914e-576e612869a5}. 

\section*{Author contributions}
 C.X.Y. and B.v.S. conducted the experiments and performed the analysis with support from V.J. A.S.I. conceived and constructed the experimental set-up. V.J and F.B designed the device. S.D.O. fabricated the device, L.E.A.S. and G.S. supplied the heterostructures. C.X.Y., B.v.S and M.V. wrote the manuscript with input from all authors. M.V. supervised the project.
 
\section*{Declarations}
M.V. and G.S. are founding advisors of Groove Quantum BV and declare equity interests. The remaining authors declare that they have no competing interests.

\section*{Methods}
The spin qubit device is fabricated on a Ge/SiGe heterostructure as detailed in Ref.\cite{AStehouwer2025ExploitingMicron-scale}. The device is cooled down in a Bluefors LD400 dilution fridge equipped with an uniaxial solenoid magnet up to \SI{3}{\tesla} in the $z$-axis at a base temperature of \SI{10}{mK}. The uniaxial magnet is operated in the driven mode except when it is turned off. The NdFeB N45 block magnet is commercially available under article ID Q-111-89-20-E,  its dimensions are 110.6 × 89 × 19.5 mm and it has an adhesive force of \SI{200}{kg}. Its magnetic field strength is probed using 3D magnetic Hall sensors integrated on a printed circuit board (ref. TLE493D-P3XX-MS2GO). 

In the experiments, we initialise the qubit by pulsing from the (0,2) singlet state to the center of the (1,1) state adiabatically to prepare a $\ket{\uparrow, \downarrow}$ state. We then perform qubit manipulation via EDSR before reading out the spin state using Pauli spin blockade readout. 
The dc voltages are set to the device via an in-house built SPI rack (\url{https://qtwork.tudelft.nl/~mtiggelman/spi-rack/chassis.html}) and ac controls are generated using a Keysight M3202A arbitrary waveform generator (AWG).

\FloatBarrier
\bibliography{references.bib}

\clearpage
\onecolumngrid

\setcounter{section}{0}
\renewcommand{\thesection}{\arabic{section}}
\section{Additional qubit data}
\label{appendix:Q3}

Here we plot the coherence times as a function of the magnet position along $x$ direction for a qubit Q3 that located on the east side of the device, see Figure-1c in the main text. This qubit demonstrates the same trend in Larmor frequency and coherence times as a function of the magnet $x$-position. Note that we cannot compare the sweet spot position in $x$-direction between Q3 and Q8 investigated in the main text due to hysteresis effect of the magnet movement (see Supplementary Note 6).  

\begin{figure*}[h!]
    \includegraphics{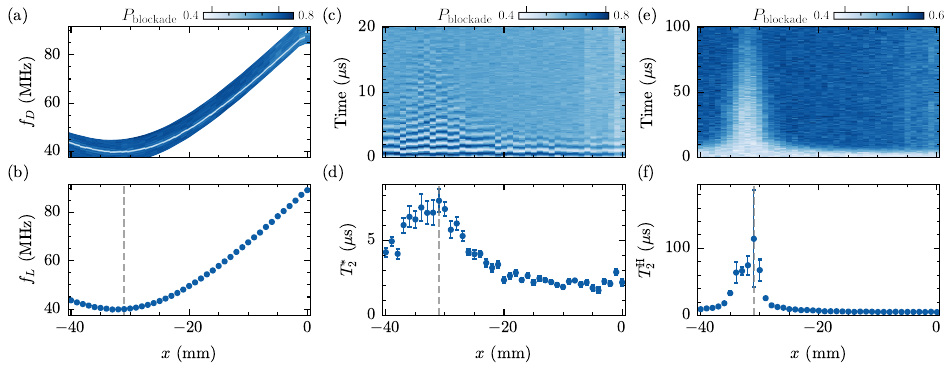}
	\caption{\textbf{a, b,} Larmor frequency $f_\mathrm{L}$ as a function of the \textit{x}-position of the permanent magnet extracted from the microwave spectroscopy. \textbf{c,d,} Dephasing times $T_2^*$ as a function of the magnet position in $x$ measured with a Ramsey sequence. \textbf{e, f,} $T_2^\mathrm{H}$ measured with a Hahn echo sequence. For both Ramsey and Hahn echo sequences, $t_\mathrm{wait}$ denotes the total free evolution time. The dashed grey lines indicate the optimal field spot at the minimum of the Larmor frequency where the magnetic field points in-plane. The error bars represent the precision of the fit.}
    \label{fig:Q3}
\end{figure*}

\section{Qubit coherence times without the external magnet}
\label{appendix:no_magnet}

\begin{figure*}[h!]
    \includegraphics{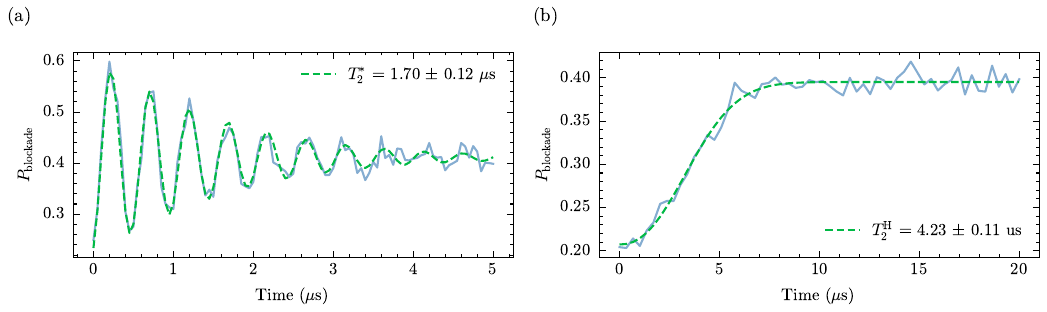}
	\caption{Coherence times of the qubit without the external magnet. The blue curves represent the measurement data, and the green dashed lines show the fits for a Ramsey sequence (\textbf{a}) and a Hahn-echo sequence (\textbf{b}).}
    \label{fig:coherence_no_magnet}
\end{figure*}

In \autoref{fig:coherence_no_magnet}, we measure the coherence times of qubit Q8 by operating with the internal magnet set to \SI{25}{mT} while the external magnet is \SI{700}{mm} away from the sample such that we assume it has no effect on the total magnetic field. We estimate that here the magnetic field points 2-3 degrees out-of-plane and the qubit coherence times are expected to be hyperfine limited \cite{AStehouwer2025ExploitingMicron-scale, John2024AControl}.

\section{Driving efficiency}
\label{appendix:efficiency}

Here we plot the Rabi drive efficiency and the dephasing time $T_2^*$ of Q8 as a function of the external magnet position in the $x$ direction. 

\begin{figure*}[h!]
    \centering
    \includegraphics[width=0.5\linewidth]{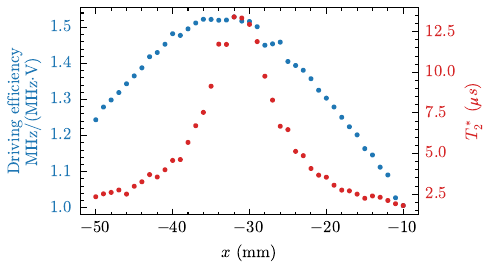}
    \caption{We define the driving efficiency as $f_\mathrm{Rabi} / (f_\mathrm{L}\cdot A)$, with $f_\mathrm{Rabi}$ being the Rabi frequency, $f_\mathrm{L}$ the Larmor frequency and $A$ the drive amplitude at the qubit level.}
    \label{fig:rabi}
\end{figure*}

\section{Randomised Benchmarking}
\label{appendix:rb}

\begin{figure}[h]
	\includegraphics{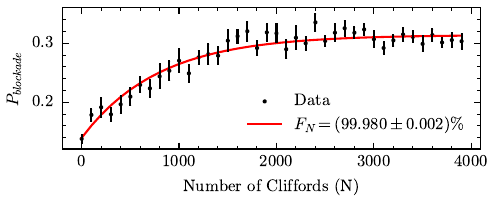}
	\centering
    \caption{Single qubit gate RB data at the in-plane magnetic field. $F_\mathrm{N}$ represents the average gate fidelity using the Clifford set detailed in \autoref{tab: rb_clifford}. 
    The error bars represent the uncertainty in the average spin probability over 20 randomisations.
    The uncertainty on the gate fidelity corresponds to fitting errors. }
    \label{fig: RB}
\end{figure}

In this work we perform single qubit RB in an identical fashion to that described in \cite{Lawrie2023SimultaneousThreshold}. In particular, we randomly sample Clifford gates, composed from the native X90 and Y90 gates according to \autoref{tab: rb_clifford}. We then append a recovery gate, taken to be the Clifford gate which is the inverse of the sequence proceeding it. We evaluate the outcomes of this sequence for increasing numbers of randomly sampled Clifford gates. We then fit a decaying exponential of the form 
\begin{equation*}
    P_\text{blockade} = A \alpha^{N_C} + B,
\end{equation*}
where $A$ represents the visibility of the system, $\alpha$ is the depolarizing parameter, $N_c$ is the number of Clifford gates in the sequence, and $B$ corresponds to the average signal of the $\left|00\right\rangle$ and $\left|01\right\rangle$ subspace (where the first index corresponds to the qubit being benchmarked) \cite{Xue2019BenchmarkingDevice}. For a single qubit, the depolarizating parameter is related to the average Clifford fidelity according to 
\begin{equation*}
    F_C = 1 - (1- \alpha) / 2.
\end{equation*}
For comparison with GST we then convert the Clifford gate fidelity, $F_C$ to a native gate fidelity $F_N$, this is achieved by dividing the Clifford infidelity by the average number of gates per Clifford, such that 
\begin{equation*}
    1 - F_N = (1 - F_C)/n_C,
\end{equation*}
where $n_C$ is the average number of native gates per Clifford, which in our case is $3.217$. Single qubit benchmark of Q3 at the hyperfine noise sweet spot is shown in \autoref{fig: RB}.

\begin{table}[h]
\centering\label{tab:rb_clifford}
\begin{tabular}{|l|l|}
\hline
\textbf{Operation} & \textbf{Pulse Sequence} \\ \hline
Pauli I (Identity) & $I$ \\ \hline
\multirow{3}{*}{Pauli $\pi$ rotations} 
    & $X_{90} X_{90}$ (X) \\
    & $Y_{90} Y_{90}$ (Y) \\
    & $Y_{90} Y_{90} X_{90} X_{90}$ (Z) \\ \hline
\multirow{8}{*}{$2\pi/3$ rotations around $\sqrt{1/3}[\pm1,\pm1,\pm1]$}
    & $X_{90} Y_{90}$ \\
    & $X_{90} Y_{90} Y_{90} Y_{90}$ \\
    & $X_{90} X_{90} X_{90} Y_{90}$ \\
    & $Y_{90} Y_{90} X_{90} Y_{90}$ \\
    & $Y_{90} X_{90}$ \\
    & $Y_{90} X_{90} X_{90} X_{90}$ \\
    & $Y_{90} Y_{90} Y_{90} X_{90}$ \\
    & $Y_{90} X_{90} Y_{90} Y_{90}$ \\ \hline
\multirow{6}{*}{$\pi/2$ rotations around $[\pm1,0,0]$, $[0,\pm1,0]$, $[0,0,\pm1]$}
    & $X_{90}$ \\
    & $X_{90} X_{90} X_{90}$ \\
    & $Y_{90}$ \\
    & $Y_{90} Y_{90} Y_{90}$ \\
    & $Y_{90} X_{90} Y_{90} Y_{90} Y_{90}$ \\
    & $Y_{90} Y_{90} Y_{90} X_{90} Y_{90}$ \\ \hline
\multirow{6}{*}{Hadamard-like $\pi$ rotations}
    & $X_{90} X_{90} Y_{90}$ \\
    & $Y_{90} X_{90} X_{90}$ \\
    & $Y_{90} Y_{90} X_{90}$ \\
    & $X_{90} Y_{90} Y_{90}$ \\
    & $Y_{90} X_{90} Y_{90}$ \\
    & $Y_{90} X_{90} X_{90} X_{90} Y_{90}$ \\ \hline
\end{tabular}
\caption{RB Clifford gates decomposed into native $X_{90}$ and $Y_{90}$ pulses \cite{Lawrie2023SimultaneousThreshold}. The gates are ordered from left to right to indicate their temporal sequence, as opposed to the operator notation where the first gate is the rightmost. } \label{tab: rb_clifford}
\end{table}

\section{Gate set tomography}
\label{appendix:gst}

\begin{figure*}
	\includegraphics{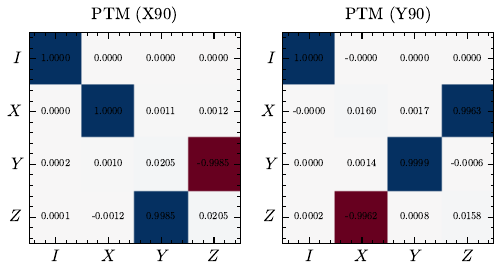}
	\caption{Pauli transfer matrix (PTM) of the X90 and Y90 gates, as constructed by pyGSTi, using the $XY$ model \cite{Nielsen2021GateTomography}.}
    \label{fig:gst}
\end{figure*}

To perform GST, we use the XY model implemented by pyGSTi \cite{Nielsen2021GateTomography}, with a max circuit length of 128 gates. We then use pyGSTi to fit the resulting data. The table of individual error metrics generated in the report is shown in Table \ref{tab: error_metrics}. The Pauli transfer matrix for the X90 and Y90 gate is shown in \autoref{fig:gst}.

\begin{table}
    \centering
    \caption{Individual gate fidelities from GST report (Gauge variant)}
    \begin{tabular}{|c|c|c|c|c|c|}
        \hline
        \textbf{Gate} &
        \begin{tabular}[c]{@{}c@{}}\textbf{Entanglement}\\ \textbf{Fidelity (\%)}\end{tabular} &
        \begin{tabular}[c]{@{}c@{}}\textbf{Avg. Gate}\\ \textbf{Fidelity (\%)}\end{tabular} &
        \begin{tabular}[c]{@{}c@{}}$1/2$ \\ \textbf{Trace Distance}\end{tabular} &
        \begin{tabular}[c]{@{}c@{}}\textbf{Non-unitary}\\ \textbf{Ent. Fidelity (\%)}\end{tabular} &
        \begin{tabular}[c]{@{}c@{}}\textbf{Non-unitary}\\ \textbf{Avg. Fidelity (\%)}\end{tabular} \\
        \hline
        X90 & $99.9232 \pm 0.025$ & $99.9488 \pm 0.0167$ & $0.010309 \pm 0.000402$ & $99.9338 \pm 0.029$ & $99.9559 \pm 0.0166$ \\
        \hline
        Y90 & $99.8080 \pm 0.073$ & $99.8762 \pm 0.0192$ & $0.008284 \pm 0.000359$ & $99.8145 \pm 0.0287$ & $99.8763 \pm 0.0191$ \\
        \hline
    \end{tabular}
    \label{tab: error_metrics}
\end{table}

\clearpage

\section{Hysteresis effect in the magnet movement}
\label{appendix:hysteresis}

\begin{figure*}
    \includegraphics{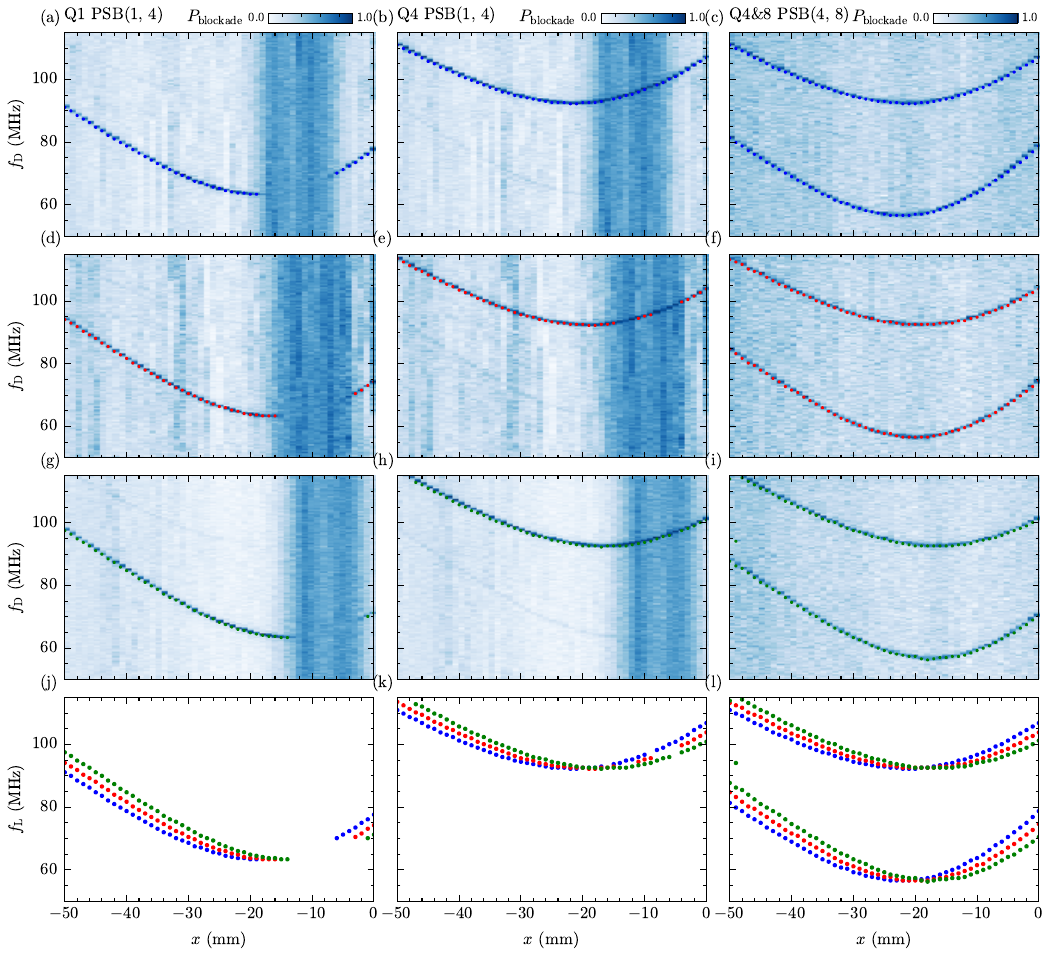}
    \caption{Hysteresis effect of the magnet movement probed over three different qubits Q1, Q4 and Q8. PSB($i,j$) indicates that the readout is performed between qubits Q$i$ and Q$j$. \textbf{a-c,} Microwave spectroscopy of qubits Q1, Q4, Q8 during the first magnet run. \textbf{d-f,} Second run \textbf{g-i,} Third run. \textbf{j-l,} Extracted Larmor frequencies for each qubit across the three runs (blue: first run, red: second run, green: third run) showing the run-to-run variation in qubit response to the external magnet position.}
    \label{fig:hysteresis}
\end{figure*}

Here, we quantify the hysteresis effect observed in the magnet motion on three different qubits Q1, Q4 and Q8. The movement of the magnet is non-linear, which introduces hysteresis when repeating runs involving multiple start-and-stop motions. From \autoref{fig:hysteresis}, we estimate an offset of \SI{2.5}{mm} per run of 51 data points. This corresponds to an approximate offset of \SI{50}{\um} per start-and-stop event during a magnet movement of \SI{1}{mm}. Given the reproducibility of the offset, it could be compensated in software by a correction of \SI{50}{\um} to each data point. 
The system can also be improved by replacing the motors with ones that include integrated calibration routines, similar to those used in 3D printers, or by incorporating a distance measurement system that provides real-time feedback on the magnet position to enable in-situ calibration of the existing setup.

\bibliography{references.bib}

\begin{thebibliography}{23}%
\makeatletter
\providecommand \@ifxundefined [1]{%
 \@ifx{#1\undefined}
}%
\providecommand \@ifnum [1]{%
 \ifnum #1\expandafter \@firstoftwo
 \else \expandafter \@secondoftwo
 \fi
}%
\providecommand \@ifx [1]{%
 \ifx #1\expandafter \@firstoftwo
 \else \expandafter \@secondoftwo
 \fi
}%
\providecommand \natexlab [1]{#1}%
\providecommand \enquote  [1]{``#1''}%
\providecommand \bibnamefont  [1]{#1}%
\providecommand \bibfnamefont [1]{#1}%
\providecommand \citenamefont [1]{#1}%
\providecommand \href@noop [0]{\@secondoftwo}%
\providecommand \href [0]{\begingroup \@sanitize@url \@href}%
\providecommand \@href[1]{\@@startlink{#1}\@@href}%
\providecommand \@@href[1]{\endgroup#1\@@endlink}%
\providecommand \@sanitize@url [0]{\catcode `\\12\catcode `\$12\catcode `\&12\catcode `\#12\catcode `\^12\catcode `\_12\catcode `\%12\relax}%
\providecommand \@@startlink[1]{}%
\providecommand \@@endlink[0]{}%
\providecommand \url  [0]{\begingroup\@sanitize@url \@url }%
\providecommand \@url [1]{\endgroup\@href {#1}{\urlprefix }}%
\providecommand \urlprefix  [0]{URL }%
\providecommand \Eprint [0]{\href }%
\providecommand \doibase [0]{https://doi.org/}%
\providecommand \selectlanguage [0]{\@gobble}%
\providecommand \bibinfo  [0]{\@secondoftwo}%
\providecommand \bibfield  [0]{\@secondoftwo}%
\providecommand \translation [1]{[#1]}%
\providecommand \BibitemOpen [0]{}%
\providecommand \bibitemStop [0]{}%
\providecommand \bibitemNoStop [0]{.\EOS\space}%
\providecommand \EOS [0]{\spacefactor3000\relax}%
\providecommand \BibitemShut  [1]{\csname bibitem#1\endcsname}%
\let\auto@bib@innerbib\@empty
\bibitem [{\citenamefont {Burkard}\ \emph {et~al.}(2023)\citenamefont {Burkard}, \citenamefont {Ladd}, \citenamefont {Pan}, \citenamefont {Nichol},\ and\ \citenamefont {Petta}}]{Burkard2023SemiconductorQubits}%
  \BibitemOpen
  \bibfield  {author} {\bibinfo {author} {\bibfnamefont {G.}~\bibnamefont {Burkard}}, \bibinfo {author} {\bibfnamefont {T.~D.}\ \bibnamefont {Ladd}}, \bibinfo {author} {\bibfnamefont {A.}~\bibnamefont {Pan}}, \bibinfo {author} {\bibfnamefont {J.~M.}\ \bibnamefont {Nichol}},\ and\ \bibinfo {author} {\bibfnamefont {J.~R.}\ \bibnamefont {Petta}},\ }\bibfield  {title} {\bibinfo {title} {{Semiconductor spin qubits}},\ }\href {https://doi.org/10.1103/REVMODPHYS.95.025003} {\bibfield  {journal} {\bibinfo  {journal} {Reviews of Modern Physics}\ }\textbf {\bibinfo {volume} {95}},\ \bibinfo {pages} {025003} (\bibinfo {year} {2023})}\BibitemShut {NoStop}%
\bibitem [{\citenamefont {Watzinger}\ \emph {et~al.}(2018)\citenamefont {Watzinger}, \citenamefont {Kuku{\v{c}}ka}, \citenamefont {Vuku{\v{s}}i{\'{c}}}, \citenamefont {Gao}, \citenamefont {Wang}, \citenamefont {Sch{\"{a}}ffler}, \citenamefont {Zhang},\ and\ \citenamefont {Katsaros}}]{Watzinger2018AQubit}%
  \BibitemOpen
  \bibfield  {author} {\bibinfo {author} {\bibfnamefont {H.}~\bibnamefont {Watzinger}}, \bibinfo {author} {\bibfnamefont {J.}~\bibnamefont {Kuku{\v{c}}ka}}, \bibinfo {author} {\bibfnamefont {L.}~\bibnamefont {Vuku{\v{s}}i{\'{c}}}}, \bibinfo {author} {\bibfnamefont {F.}~\bibnamefont {Gao}}, \bibinfo {author} {\bibfnamefont {T.}~\bibnamefont {Wang}}, \bibinfo {author} {\bibfnamefont {F.}~\bibnamefont {Sch{\"{a}}ffler}}, \bibinfo {author} {\bibfnamefont {J.~J.}\ \bibnamefont {Zhang}},\ and\ \bibinfo {author} {\bibfnamefont {G.}~\bibnamefont {Katsaros}},\ }\bibfield  {title} {\bibinfo {title} {{A germanium hole spin qubit}},\ }\href {https://doi.org/10.1038/s41467-018-06418-4} {\bibfield  {journal} {\bibinfo  {journal} {Nature Communications}\ }\textbf {\bibinfo {volume} {9}},\ \bibinfo {pages} {1} (\bibinfo {year} {2018})}\BibitemShut {NoStop}%
\bibitem [{\citenamefont {Hendrickx}\ \emph {et~al.}(2021)\citenamefont {Hendrickx}, \citenamefont {Lawrie}, \citenamefont {Russ}, \citenamefont {van Riggelen}, \citenamefont {de~Snoo}, \citenamefont {Schouten}, \citenamefont {Sammak}, \citenamefont {Scappucci},\ and\ \citenamefont {Veldhorst}}]{Hendrickx2021AProcessor}%
  \BibitemOpen
  \bibfield  {author} {\bibinfo {author} {\bibfnamefont {N.~W.}\ \bibnamefont {Hendrickx}}, \bibinfo {author} {\bibfnamefont {W.~I.}\ \bibnamefont {Lawrie}}, \bibinfo {author} {\bibfnamefont {M.}~\bibnamefont {Russ}}, \bibinfo {author} {\bibfnamefont {F.}~\bibnamefont {van Riggelen}}, \bibinfo {author} {\bibfnamefont {S.~L.}\ \bibnamefont {de~Snoo}}, \bibinfo {author} {\bibfnamefont {R.~N.}\ \bibnamefont {Schouten}}, \bibinfo {author} {\bibfnamefont {A.}~\bibnamefont {Sammak}}, \bibinfo {author} {\bibfnamefont {G.}~\bibnamefont {Scappucci}},\ and\ \bibinfo {author} {\bibfnamefont {M.}~\bibnamefont {Veldhorst}},\ }\bibfield  {title} {\bibinfo {title} {{A four-qubit germanium quantum processor}},\ }\href {https://doi.org/10.1038/s41586-021-03332-6} {\bibfield  {journal} {\bibinfo  {journal} {Nature}\ }\textbf {\bibinfo {volume} {591}},\ \bibinfo {pages} {580} (\bibinfo {year} {2021})}\BibitemShut {NoStop}%
\bibitem [{\citenamefont {Wang}\ \emph {et~al.}(2024)\citenamefont {Wang}, \citenamefont {John}, \citenamefont {Tidjani}, \citenamefont {Yu}, \citenamefont {Ivlev}, \citenamefont {D{\'{e}}prez}, \citenamefont {van Riggelen-Doelman}, \citenamefont {Woods}, \citenamefont {Hendrickx}, \citenamefont {Lawrie}, \citenamefont {Stehouwer}, \citenamefont {Oosterhout}, \citenamefont {Sammak}, \citenamefont {Friesen}, \citenamefont {Scappucci}, \citenamefont {de~Snoo}, \citenamefont {Rimbach-Russ}, \citenamefont {Borsoi},\ and\ \citenamefont {Veldhorst}}]{Wang2024OperatingSpins}%
  \BibitemOpen
  \bibfield  {author} {\bibinfo {author} {\bibfnamefont {C.~A.}\ \bibnamefont {Wang}}, \bibinfo {author} {\bibfnamefont {V.}~\bibnamefont {John}}, \bibinfo {author} {\bibfnamefont {H.}~\bibnamefont {Tidjani}}, \bibinfo {author} {\bibfnamefont {C.~X.}\ \bibnamefont {Yu}}, \bibinfo {author} {\bibfnamefont {A.~S.}\ \bibnamefont {Ivlev}}, \bibinfo {author} {\bibfnamefont {C.}~\bibnamefont {D{\'{e}}prez}}, \bibinfo {author} {\bibfnamefont {F.}~\bibnamefont {van Riggelen-Doelman}}, \bibinfo {author} {\bibfnamefont {B.~D.}\ \bibnamefont {Woods}}, \bibinfo {author} {\bibfnamefont {N.~W.}\ \bibnamefont {Hendrickx}}, \bibinfo {author} {\bibfnamefont {W.~I.}\ \bibnamefont {Lawrie}}, \bibinfo {author} {\bibfnamefont {L.~E.}\ \bibnamefont {Stehouwer}}, \bibinfo {author} {\bibfnamefont {S.~D.}\ \bibnamefont {Oosterhout}}, \bibinfo {author} {\bibfnamefont {A.}~\bibnamefont {Sammak}}, \bibinfo {author} {\bibfnamefont {M.}~\bibnamefont {Friesen}}, \bibinfo {author} {\bibfnamefont {G.}~\bibnamefont {Scappucci}}, \bibinfo
  {author} {\bibfnamefont {S.~L.}\ \bibnamefont {de~Snoo}}, \bibinfo {author} {\bibfnamefont {M.}~\bibnamefont {Rimbach-Russ}}, \bibinfo {author} {\bibfnamefont {F.}~\bibnamefont {Borsoi}},\ and\ \bibinfo {author} {\bibfnamefont {M.}~\bibnamefont {Veldhorst}},\ }\bibfield  {title} {\bibinfo {title} {{Operating semiconductor quantum processors with hopping spins}},\ }\href {https://doi.org/10.1126/SCIENCE.ADO5915} {\bibfield  {journal} {\bibinfo  {journal} {Science}\ }\textbf {\bibinfo {volume} {385}},\ \bibinfo {pages} {447} (\bibinfo {year} {2024})}\BibitemShut {NoStop}%
\bibitem [{\citenamefont {Scappucci}\ \emph {et~al.}(2020)\citenamefont {Scappucci}, \citenamefont {Kloeffel}, \citenamefont {Zwanenburg}, \citenamefont {Loss}, \citenamefont {Myronov}, \citenamefont {Zhang}, \citenamefont {De~Franceschi}, \citenamefont {Katsaros},\ and\ \citenamefont {Veldhorst}}]{Scappucci2020TheRoute}%
  \BibitemOpen
  \bibfield  {author} {\bibinfo {author} {\bibfnamefont {G.}~\bibnamefont {Scappucci}}, \bibinfo {author} {\bibfnamefont {C.}~\bibnamefont {Kloeffel}}, \bibinfo {author} {\bibfnamefont {F.~A.}\ \bibnamefont {Zwanenburg}}, \bibinfo {author} {\bibfnamefont {D.}~\bibnamefont {Loss}}, \bibinfo {author} {\bibfnamefont {M.}~\bibnamefont {Myronov}}, \bibinfo {author} {\bibfnamefont {J.~J.}\ \bibnamefont {Zhang}}, \bibinfo {author} {\bibfnamefont {S.}~\bibnamefont {De~Franceschi}}, \bibinfo {author} {\bibfnamefont {G.}~\bibnamefont {Katsaros}},\ and\ \bibinfo {author} {\bibfnamefont {M.}~\bibnamefont {Veldhorst}},\ }\bibfield  {title} {\bibinfo {title} {{The germanium quantum information route}},\ }\href {https://doi.org/10.1038/s41578-020-00262-z} {\bibfield  {journal} {\bibinfo  {journal} {Nature Reviews Materials}\ }\textbf {\bibinfo {volume} {6}},\ \bibinfo {pages} {926} (\bibinfo {year} {2020})}\BibitemShut {NoStop}%
\bibitem [{\citenamefont {Hendrickx}\ \emph {et~al.}(2024)\citenamefont {Hendrickx}, \citenamefont {Massai}, \citenamefont {Mergenthaler}, \citenamefont {Schupp}, \citenamefont {Paredes}, \citenamefont {Bedell}, \citenamefont {Salis},\ and\ \citenamefont {Fuhrer}}]{Hendrickx2024Sweet-spotSensitivity}%
  \BibitemOpen
  \bibfield  {author} {\bibinfo {author} {\bibfnamefont {N.~W.}\ \bibnamefont {Hendrickx}}, \bibinfo {author} {\bibfnamefont {L.}~\bibnamefont {Massai}}, \bibinfo {author} {\bibfnamefont {M.}~\bibnamefont {Mergenthaler}}, \bibinfo {author} {\bibfnamefont {F.~J.}\ \bibnamefont {Schupp}}, \bibinfo {author} {\bibfnamefont {S.}~\bibnamefont {Paredes}}, \bibinfo {author} {\bibfnamefont {S.~W.}\ \bibnamefont {Bedell}}, \bibinfo {author} {\bibfnamefont {G.}~\bibnamefont {Salis}},\ and\ \bibinfo {author} {\bibfnamefont {A.}~\bibnamefont {Fuhrer}},\ }\bibfield  {title} {\bibinfo {title} {{Sweet-spot operation of a germanium hole spin qubit with highly anisotropic noise sensitivity}},\ }\href {https://doi.org/10.1038/s41563-024-01857-5} {\bibfield  {journal} {\bibinfo  {journal} {Nature Materials}\ }\textbf {\bibinfo {volume} {23}},\ \bibinfo {pages} {920} (\bibinfo {year} {2024})}\BibitemShut {NoStop}%
\bibitem [{\citenamefont {Mauro}\ \emph {et~al.}(2024)\citenamefont {Mauro}, \citenamefont {Rodr{\'{i}}guez-Mena}, \citenamefont {Bassi}, \citenamefont {Schmitt},\ and\ \citenamefont {Niquet}}]{Mauro2024GeometryQubits}%
  \BibitemOpen
  \bibfield  {author} {\bibinfo {author} {\bibfnamefont {L.}~\bibnamefont {Mauro}}, \bibinfo {author} {\bibfnamefont {E.~A.}\ \bibnamefont {Rodr{\'{i}}guez-Mena}}, \bibinfo {author} {\bibfnamefont {M.}~\bibnamefont {Bassi}}, \bibinfo {author} {\bibfnamefont {V.}~\bibnamefont {Schmitt}},\ and\ \bibinfo {author} {\bibfnamefont {Y.~M.}\ \bibnamefont {Niquet}},\ }\bibfield  {title} {\bibinfo {title} {{Geometry of the dephasing sweet spots of spin-orbit qubits}},\ }\href {https://doi.org/10.1103/PHYSREVB.109.155406} {\bibfield  {journal} {\bibinfo  {journal} {Physical Review B}\ }\textbf {\bibinfo {volume} {109}},\ \bibinfo {pages} {155406} (\bibinfo {year} {2024})}\BibitemShut {NoStop}%
\bibitem [{\citenamefont {Saez-Mollejo}\ \emph {et~al.}(2025)\citenamefont {Saez-Mollejo}, \citenamefont {Jirovec}, \citenamefont {Schell}, \citenamefont {Kukucka}, \citenamefont {Calcaterra}, \citenamefont {Chrastina}, \citenamefont {Isella}, \citenamefont {Rimbach-Russ}, \citenamefont {Bosco},\ and\ \citenamefont {Katsaros}}]{Saez-Mollejo2025ExchangeQubits}%
  \BibitemOpen
  \bibfield  {author} {\bibinfo {author} {\bibfnamefont {J.}~\bibnamefont {Saez-Mollejo}}, \bibinfo {author} {\bibfnamefont {D.}~\bibnamefont {Jirovec}}, \bibinfo {author} {\bibfnamefont {Y.}~\bibnamefont {Schell}}, \bibinfo {author} {\bibfnamefont {J.}~\bibnamefont {Kukucka}}, \bibinfo {author} {\bibfnamefont {S.}~\bibnamefont {Calcaterra}}, \bibinfo {author} {\bibfnamefont {D.}~\bibnamefont {Chrastina}}, \bibinfo {author} {\bibfnamefont {G.}~\bibnamefont {Isella}}, \bibinfo {author} {\bibfnamefont {M.}~\bibnamefont {Rimbach-Russ}}, \bibinfo {author} {\bibfnamefont {S.}~\bibnamefont {Bosco}},\ and\ \bibinfo {author} {\bibfnamefont {G.}~\bibnamefont {Katsaros}},\ }\bibfield  {title} {\bibinfo {title} {{Exchange anisotropies in microwave-driven singlet-triplet qubits}},\ }\href {https://doi.org/10.1038/s41467-025-58969-y} {\bibfield  {journal} {\bibinfo  {journal} {Nature Communications}\ }\textbf {\bibinfo {volume} {16}},\ \bibinfo {pages} {1} (\bibinfo {year} {2025})}\BibitemShut {NoStop}%
\bibitem [{\citenamefont {John}\ \emph {et~al.}(2024)\citenamefont {John}, \citenamefont {Yu}, \citenamefont {Van~Straaten}, \citenamefont {Rodr{\'{i}}guez-Mena}, \citenamefont {Rodr{\'{i}}guez}, \citenamefont {Oosterhout}, \citenamefont {Stehouwer}, \citenamefont {Scappucci}, \citenamefont {Bosco}, \citenamefont {Rimbach-Russ}, \citenamefont {Niquet}, \citenamefont {Borsoi},\ and\ \citenamefont {Veldhorst}}]{John2024AControl}%
  \BibitemOpen
  \bibfield  {author} {\bibinfo {author} {\bibfnamefont {V.}~\bibnamefont {John}}, \bibinfo {author} {\bibfnamefont {C.~X.}\ \bibnamefont {Yu}}, \bibinfo {author} {\bibfnamefont {B.}~\bibnamefont {Van~Straaten}}, \bibinfo {author} {\bibfnamefont {E.~A.}\ \bibnamefont {Rodr{\'{i}}guez-Mena}}, \bibinfo {author} {\bibfnamefont {M.}~\bibnamefont {Rodr{\'{i}}guez}}, \bibinfo {author} {\bibfnamefont {S.}~\bibnamefont {Oosterhout}}, \bibinfo {author} {\bibfnamefont {L.~E.~A.}\ \bibnamefont {Stehouwer}}, \bibinfo {author} {\bibfnamefont {G.}~\bibnamefont {Scappucci}}, \bibinfo {author} {\bibfnamefont {S.}~\bibnamefont {Bosco}}, \bibinfo {author} {\bibfnamefont {M.}~\bibnamefont {Rimbach-Russ}}, \bibinfo {author} {\bibfnamefont {Y.-M.}\ \bibnamefont {Niquet}}, \bibinfo {author} {\bibfnamefont {F.}~\bibnamefont {Borsoi}},\ and\ \bibinfo {author} {\bibfnamefont {M.}~\bibnamefont {Veldhorst}},\ }\bibfield  {title} {\bibinfo {title} {{A two-dimensional 10-qubit array in germanium with robust and localised qubit
  control}},\ }\href {https://doi.org/10.48550/arXiv.2412.16044} {\bibfield  {journal} {\bibinfo  {journal} {arXiv}\ }\textbf {\bibinfo {volume} {2412.16044}} (\bibinfo {year} {2024})}\BibitemShut {NoStop}%
\bibitem [{\citenamefont {Bassi}\ \emph {et~al.}(2024)\citenamefont {Bassi}, \citenamefont {Rodr{\'{i}}guez-Mena}, \citenamefont {Brun}, \citenamefont {Zihlmann}, \citenamefont {Nguyen}, \citenamefont {Champain}, \citenamefont {Abadillo-Uriel}, \citenamefont {Bertrand}, \citenamefont {Niebojewski}, \citenamefont {Maurand}, \citenamefont {Niquet}, \citenamefont {Jehl}, \citenamefont {De~Franceschi},\ and\ \citenamefont {Schmitt}}]{Bassi2024OptimalQubits}%
  \BibitemOpen
  \bibfield  {author} {\bibinfo {author} {\bibfnamefont {M.}~\bibnamefont {Bassi}}, \bibinfo {author} {\bibfnamefont {E.~A.}\ \bibnamefont {Rodr{\'{i}}guez-Mena}}, \bibinfo {author} {\bibfnamefont {B.}~\bibnamefont {Brun}}, \bibinfo {author} {\bibfnamefont {S.}~\bibnamefont {Zihlmann}}, \bibinfo {author} {\bibfnamefont {T.}~\bibnamefont {Nguyen}}, \bibinfo {author} {\bibfnamefont {V.}~\bibnamefont {Champain}}, \bibinfo {author} {\bibfnamefont {J.~C.}\ \bibnamefont {Abadillo-Uriel}}, \bibinfo {author} {\bibfnamefont {B.}~\bibnamefont {Bertrand}}, \bibinfo {author} {\bibfnamefont {H.}~\bibnamefont {Niebojewski}}, \bibinfo {author} {\bibfnamefont {R.}~\bibnamefont {Maurand}}, \bibinfo {author} {\bibfnamefont {Y.-M.}\ \bibnamefont {Niquet}}, \bibinfo {author} {\bibfnamefont {X.}~\bibnamefont {Jehl}}, \bibinfo {author} {\bibfnamefont {S.}~\bibnamefont {De~Franceschi}},\ and\ \bibinfo {author} {\bibfnamefont {V.}~\bibnamefont {Schmitt}},\ }\bibfield  {title} {\bibinfo {title} {{Optimal operation of hole spin
  qubits}},\ }\href {https://doi.org/10.48550/arXiv.2412.13069} {\bibfield  {journal} {\bibinfo  {journal} {arXiv}\ }\textbf {\bibinfo {volume} {2412.13069}} (\bibinfo {year} {2024})}\BibitemShut {NoStop}%
\bibitem [{\citenamefont {Da~Prato}\ \emph {et~al.}(2025)\citenamefont {Da~Prato}, \citenamefont {Yu}, \citenamefont {Bode},\ and\ \citenamefont {Gr{\"{o}}blacher}}]{DaPrato2025Step-by-stepMagnet}%
  \BibitemOpen
  \bibfield  {author} {\bibinfo {author} {\bibfnamefont {G.}~\bibnamefont {Da~Prato}}, \bibinfo {author} {\bibfnamefont {Y.}~\bibnamefont {Yu}}, \bibinfo {author} {\bibfnamefont {R.}~\bibnamefont {Bode}},\ and\ \bibinfo {author} {\bibfnamefont {S.}~\bibnamefont {Gr{\"{o}}blacher}},\ }\bibfield  {title} {\bibinfo {title} {{Step-by-step design guide of a cryogenic three-axis vector magnet}},\ }\href {https://doi.org/10.1063/5.0270187} {\bibfield  {journal} {\bibinfo  {journal} {arXiv}\ }\textbf {\bibinfo {volume} {2503.05459}} (\bibinfo {year} {2025})}\BibitemShut {NoStop}%
\bibitem [{\citenamefont {A~Stehouwer}\ \emph {et~al.}(2025)\citenamefont {A~Stehouwer}, \citenamefont {Yu}, \citenamefont {van Straaten}, \citenamefont {Tosato}, \citenamefont {John}, \citenamefont {Degli~Esposti}, \citenamefont {Elsayed}, \citenamefont {Costa}, \citenamefont {Oosterhout}, \citenamefont {Hendrickx}, \citenamefont {Veldhorst}, \citenamefont {Borsoi},\ and\ \citenamefont {Scappucci}}]{AStehouwer2025ExploitingMicron-scale}%
  \BibitemOpen
  \bibfield  {author} {\bibinfo {author} {\bibfnamefont {L.~E.}\ \bibnamefont {A~Stehouwer}}, \bibinfo {author} {\bibfnamefont {C.~X.}\ \bibnamefont {Yu}}, \bibinfo {author} {\bibfnamefont {B.}~\bibnamefont {van Straaten}}, \bibinfo {author} {\bibfnamefont {A.}~\bibnamefont {Tosato}}, \bibinfo {author} {\bibfnamefont {V.}~\bibnamefont {John}}, \bibinfo {author} {\bibfnamefont {D.}~\bibnamefont {Degli~Esposti}}, \bibinfo {author} {\bibfnamefont {A.}~\bibnamefont {Elsayed}}, \bibinfo {author} {\bibfnamefont {D.}~\bibnamefont {Costa}}, \bibinfo {author} {\bibfnamefont {S.~D.}\ \bibnamefont {Oosterhout}}, \bibinfo {author} {\bibfnamefont {N.~W.}\ \bibnamefont {Hendrickx}}, \bibinfo {author} {\bibfnamefont {M.}~\bibnamefont {Veldhorst}}, \bibinfo {author} {\bibfnamefont {F.}~\bibnamefont {Borsoi}},\ and\ \bibinfo {author} {\bibfnamefont {G.}~\bibnamefont {Scappucci}},\ }\bibfield  {title} {\bibinfo {title} {{Exploiting epitaxial strained germanium for scaling low noise spin qubits at the micron-scale}},\ }\href
  {https://doi.org/10.48550/arXiv.2411.11526} {\bibfield  {journal} {\bibinfo  {journal} {arXiv}\ }\textbf {\bibinfo {volume} {2411.11526}} (\bibinfo {year} {2025})}\BibitemShut {NoStop}%
\bibitem [{\citenamefont {Jirovec}\ \emph {et~al.}(2022)\citenamefont {Jirovec}, \citenamefont {Mutter}, \citenamefont {Hofmann}, \citenamefont {Crippa}, \citenamefont {Rychetsky}, \citenamefont {Craig}, \citenamefont {Kukucka}, \citenamefont {Martins}, \citenamefont {Ballabio}, \citenamefont {Ares}, \citenamefont {Chrastina}, \citenamefont {Isella}, \citenamefont {Burkard},\ and\ \citenamefont {Katsaros}}]{Jirovec2022DynamicsDifferences}%
  \BibitemOpen
  \bibfield  {author} {\bibinfo {author} {\bibfnamefont {D.}~\bibnamefont {Jirovec}}, \bibinfo {author} {\bibfnamefont {P.~M.}\ \bibnamefont {Mutter}}, \bibinfo {author} {\bibfnamefont {A.}~\bibnamefont {Hofmann}}, \bibinfo {author} {\bibfnamefont {A.}~\bibnamefont {Crippa}}, \bibinfo {author} {\bibfnamefont {M.}~\bibnamefont {Rychetsky}}, \bibinfo {author} {\bibfnamefont {D.~L.}\ \bibnamefont {Craig}}, \bibinfo {author} {\bibfnamefont {J.}~\bibnamefont {Kukucka}}, \bibinfo {author} {\bibfnamefont {F.}~\bibnamefont {Martins}}, \bibinfo {author} {\bibfnamefont {A.}~\bibnamefont {Ballabio}}, \bibinfo {author} {\bibfnamefont {N.}~\bibnamefont {Ares}}, \bibinfo {author} {\bibfnamefont {D.}~\bibnamefont {Chrastina}}, \bibinfo {author} {\bibfnamefont {G.}~\bibnamefont {Isella}}, \bibinfo {author} {\bibfnamefont {G.}~\bibnamefont {Burkard}},\ and\ \bibinfo {author} {\bibfnamefont {G.}~\bibnamefont {Katsaros}},\ }\bibfield  {title} {\bibinfo {title} {{Dynamics of Hole Singlet-Triplet Qubits with Large g -Factor
  Differences}},\ }\href {https://doi.org/https://doi.org/10.1103/PhysRevLett.128.126803} {\bibfield  {journal} {\bibinfo  {journal} {Physical Review Letters}\ }\textbf {\bibinfo {volume} {128}},\ \bibinfo {pages} {126803} (\bibinfo {year} {2022})}\BibitemShut {NoStop}%
\bibitem [{\citenamefont {Fischer}\ \emph {et~al.}(2008)\citenamefont {Fischer}, \citenamefont {Coish}, \citenamefont {Bulaev},\ and\ \citenamefont {Loss}}]{Fischer2008SpinDot}%
  \BibitemOpen
  \bibfield  {author} {\bibinfo {author} {\bibfnamefont {J.}~\bibnamefont {Fischer}}, \bibinfo {author} {\bibfnamefont {W.~A.}\ \bibnamefont {Coish}}, \bibinfo {author} {\bibfnamefont {D.~V.}\ \bibnamefont {Bulaev}},\ and\ \bibinfo {author} {\bibfnamefont {D.}~\bibnamefont {Loss}},\ }\bibfield  {title} {\bibinfo {title} {{Spin decoherence of a heavy hole coupled to nuclear spins in a quantum dot}},\ }\href {https://doi.org/10.1103/PHYSREVB.78.155329} {\bibfield  {journal} {\bibinfo  {journal} {Physical Review B - Condensed Matter and Materials Physics}\ }\textbf {\bibinfo {volume} {78}},\ \bibinfo {pages} {155329} (\bibinfo {year} {2008})}\BibitemShut {NoStop}%
\bibitem [{\citenamefont {Lawrie}\ \emph {et~al.}(2023)\citenamefont {Lawrie}, \citenamefont {Rimbach-Russ}, \citenamefont {Riggelen}, \citenamefont {Hendrickx}, \citenamefont {Snoo}, \citenamefont {Sammak}, \citenamefont {Scappucci}, \citenamefont {Helsen},\ and\ \citenamefont {Veldhorst}}]{Lawrie2023SimultaneousThreshold}%
  \BibitemOpen
  \bibfield  {author} {\bibinfo {author} {\bibfnamefont {W.~I.}\ \bibnamefont {Lawrie}}, \bibinfo {author} {\bibfnamefont {M.}~\bibnamefont {Rimbach-Russ}}, \bibinfo {author} {\bibfnamefont {F.~v.}\ \bibnamefont {Riggelen}}, \bibinfo {author} {\bibfnamefont {N.~W.}\ \bibnamefont {Hendrickx}}, \bibinfo {author} {\bibfnamefont {S.~L.}\ \bibnamefont {Snoo}}, \bibinfo {author} {\bibfnamefont {A.}~\bibnamefont {Sammak}}, \bibinfo {author} {\bibfnamefont {G.}~\bibnamefont {Scappucci}}, \bibinfo {author} {\bibfnamefont {J.}~\bibnamefont {Helsen}},\ and\ \bibinfo {author} {\bibfnamefont {M.}~\bibnamefont {Veldhorst}},\ }\bibfield  {title} {\bibinfo {title} {{Simultaneous single-qubit driving of semiconductor spin qubits at the fault-tolerant threshold}},\ }\href {https://doi.org/10.1038/s41467-023-39334-3} {\bibfield  {journal} {\bibinfo  {journal} {Nature Communications}\ }\textbf {\bibinfo {volume} {14}},\ \bibinfo {pages} {1} (\bibinfo {year} {2023})}\BibitemShut {NoStop}%
\bibitem [{\citenamefont {Dehollain}\ \emph {et~al.}(2016)\citenamefont {Dehollain}, \citenamefont {Muhonen}, \citenamefont {Blume-Kohout}, \citenamefont {Rudinger}, \citenamefont {Gamble}, \citenamefont {Nielsen}, \citenamefont {Laucht}, \citenamefont {Simmons}, \citenamefont {Kalra}, \citenamefont {Dzurak},\ and\ \citenamefont {Morello}}]{Dehollain2016OptimizationTomography}%
  \BibitemOpen
  \bibfield  {author} {\bibinfo {author} {\bibfnamefont {J.~P.}\ \bibnamefont {Dehollain}}, \bibinfo {author} {\bibfnamefont {J.~T.}\ \bibnamefont {Muhonen}}, \bibinfo {author} {\bibfnamefont {R.}~\bibnamefont {Blume-Kohout}}, \bibinfo {author} {\bibfnamefont {K.~M.}\ \bibnamefont {Rudinger}}, \bibinfo {author} {\bibfnamefont {J.~K.}\ \bibnamefont {Gamble}}, \bibinfo {author} {\bibfnamefont {E.}~\bibnamefont {Nielsen}}, \bibinfo {author} {\bibfnamefont {A.}~\bibnamefont {Laucht}}, \bibinfo {author} {\bibfnamefont {S.}~\bibnamefont {Simmons}}, \bibinfo {author} {\bibfnamefont {R.}~\bibnamefont {Kalra}}, \bibinfo {author} {\bibfnamefont {A.~S.}\ \bibnamefont {Dzurak}},\ and\ \bibinfo {author} {\bibfnamefont {A.}~\bibnamefont {Morello}},\ }\bibfield  {title} {\bibinfo {title} {{Optimization of a solid-state electron spin qubit using gate set tomography}},\ }\href {https://doi.org/10.1088/1367-2630/18/10/103018} {\bibfield  {journal} {\bibinfo  {journal} {New Journal of Physics}\ }\textbf {\bibinfo {volume}
  {18}},\ \bibinfo {pages} {103018} (\bibinfo {year} {2016})}\BibitemShut {NoStop}%
\bibitem [{\citenamefont {Mohseni}\ \emph {et~al.}(2024)\citenamefont {Mohseni}, \citenamefont {Scherer}, \citenamefont {Johnson}, \citenamefont {Wertheim}, \citenamefont {Otten}, \citenamefont {Aadit}, \citenamefont {Alexeev}, \citenamefont {Bresniker}, \citenamefont {Camsari}, \citenamefont {Chapman}, \citenamefont {Chatterjee}, \citenamefont {Dagnew}, \citenamefont {Esposito}, \citenamefont {Fahim}, \citenamefont {Fiorentino}, \citenamefont {Gajjar}, \citenamefont {Khalid}, \citenamefont {Kong}, \citenamefont {Kulchytskyy}, \citenamefont {Kyoseva}, \citenamefont {Li}, \citenamefont {Lott}, \citenamefont {Markov}, \citenamefont {Mcdermott}, \citenamefont {Pedretti}, \citenamefont {Rao}, \citenamefont {Rieffel}, \citenamefont {Silva}, \citenamefont {Sorebo}, \citenamefont {Spentzouris}, \citenamefont {Steiner}, \citenamefont {Torosov}, \citenamefont {Venturelli}, \citenamefont {Visser}, \citenamefont {Webb}, \citenamefont {Zhan}, \citenamefont {Cohen}, \citenamefont {Ronagh}, \citenamefont {Ho},
  \citenamefont {Beausoleil},\ and\ \citenamefont {Martinis}}]{Mohseni2024HowQubits}%
  \BibitemOpen
  \bibfield  {author} {\bibinfo {author} {\bibfnamefont {M.}~\bibnamefont {Mohseni}}, \bibinfo {author} {\bibfnamefont {A.}~\bibnamefont {Scherer}}, \bibinfo {author} {\bibfnamefont {K.~G.}\ \bibnamefont {Johnson}}, \bibinfo {author} {\bibfnamefont {O.}~\bibnamefont {Wertheim}}, \bibinfo {author} {\bibfnamefont {M.}~\bibnamefont {Otten}}, \bibinfo {author} {\bibfnamefont {N.~A.}\ \bibnamefont {Aadit}}, \bibinfo {author} {\bibfnamefont {Y.}~\bibnamefont {Alexeev}}, \bibinfo {author} {\bibfnamefont {K.~M.}\ \bibnamefont {Bresniker}}, \bibinfo {author} {\bibfnamefont {K.~Y.}\ \bibnamefont {Camsari}}, \bibinfo {author} {\bibfnamefont {B.}~\bibnamefont {Chapman}}, \bibinfo {author} {\bibfnamefont {S.}~\bibnamefont {Chatterjee}}, \bibinfo {author} {\bibfnamefont {G.~A.}\ \bibnamefont {Dagnew}}, \bibinfo {author} {\bibfnamefont {A.}~\bibnamefont {Esposito}}, \bibinfo {author} {\bibfnamefont {F.}~\bibnamefont {Fahim}}, \bibinfo {author} {\bibfnamefont {M.}~\bibnamefont {Fiorentino}}, \bibinfo {author} {\bibfnamefont
  {A.}~\bibnamefont {Gajjar}}, \bibinfo {author} {\bibfnamefont {A.}~\bibnamefont {Khalid}}, \bibinfo {author} {\bibfnamefont {X.}~\bibnamefont {Kong}}, \bibinfo {author} {\bibfnamefont {B.}~\bibnamefont {Kulchytskyy}}, \bibinfo {author} {\bibfnamefont {E.}~\bibnamefont {Kyoseva}}, \bibinfo {author} {\bibfnamefont {R.}~\bibnamefont {Li}}, \bibinfo {author} {\bibfnamefont {P.~A.}\ \bibnamefont {Lott}}, \bibinfo {author} {\bibfnamefont {I.~L.}\ \bibnamefont {Markov}}, \bibinfo {author} {\bibfnamefont {R.~F.}\ \bibnamefont {Mcdermott}}, \bibinfo {author} {\bibfnamefont {G.}~\bibnamefont {Pedretti}}, \bibinfo {author} {\bibfnamefont {P.}~\bibnamefont {Rao}}, \bibinfo {author} {\bibfnamefont {E.}~\bibnamefont {Rieffel}}, \bibinfo {author} {\bibfnamefont {A.}~\bibnamefont {Silva}}, \bibinfo {author} {\bibfnamefont {J.}~\bibnamefont {Sorebo}}, \bibinfo {author} {\bibfnamefont {P.}~\bibnamefont {Spentzouris}}, \bibinfo {author} {\bibfnamefont {Z.}~\bibnamefont {Steiner}}, \bibinfo {author} {\bibfnamefont
  {B.}~\bibnamefont {Torosov}}, \bibinfo {author} {\bibfnamefont {D.}~\bibnamefont {Venturelli}}, \bibinfo {author} {\bibfnamefont {R.~J.}\ \bibnamefont {Visser}}, \bibinfo {author} {\bibfnamefont {Z.}~\bibnamefont {Webb}}, \bibinfo {author} {\bibfnamefont {X.}~\bibnamefont {Zhan}}, \bibinfo {author} {\bibfnamefont {Y.}~\bibnamefont {Cohen}}, \bibinfo {author} {\bibfnamefont {P.}~\bibnamefont {Ronagh}}, \bibinfo {author} {\bibfnamefont {A.}~\bibnamefont {Ho}}, \bibinfo {author} {\bibfnamefont {R.~G.}\ \bibnamefont {Beausoleil}},\ and\ \bibinfo {author} {\bibfnamefont {J.~M.}\ \bibnamefont {Martinis}},\ }\bibfield  {title} {\bibinfo {title} {{How to Build a Quantum Supercomputer: Scaling from Hundreds to Millions of Qubits}},\ }\href {https://doi.org/10.48550/arXiv.2411.10406} {\bibfield  {journal} {\bibinfo  {journal} {arXiv}\ }\textbf {\bibinfo {volume} {2411.10406}} (\bibinfo {year} {2024})}\BibitemShut {NoStop}%
\bibitem [{\citenamefont {Adambukulam}\ \emph {et~al.}(2021)\citenamefont {Adambukulam}, \citenamefont {Sewani}, \citenamefont {Stemp}, \citenamefont {Asaad}, \citenamefont {M{\c{a}}dzik}, \citenamefont {Morello},\ and\ \citenamefont {Laucht}}]{Adambukulam2021AnTemperatures}%
  \BibitemOpen
  \bibfield  {author} {\bibinfo {author} {\bibfnamefont {C.}~\bibnamefont {Adambukulam}}, \bibinfo {author} {\bibfnamefont {V.~K.}\ \bibnamefont {Sewani}}, \bibinfo {author} {\bibfnamefont {H.~G.}\ \bibnamefont {Stemp}}, \bibinfo {author} {\bibfnamefont {S.}~\bibnamefont {Asaad}}, \bibinfo {author} {\bibfnamefont {M.~T.}\ \bibnamefont {M{\c{a}}dzik}}, \bibinfo {author} {\bibfnamefont {A.}~\bibnamefont {Morello}},\ and\ \bibinfo {author} {\bibfnamefont {A.}~\bibnamefont {Laucht}},\ }\bibfield  {title} {\bibinfo {title} {{An ultra-stable 1.5 T permanent magnet assembly for qubit experiments at cryogenic temperatures}},\ }\bibfield  {journal} {\bibinfo  {journal} {Review of Scientific Instruments}\ }\textbf {\bibinfo {volume} {92}},\ \href {https://doi.org/10.1063/5.0055318} {10.1063/5.0055318} (\bibinfo {year} {2021})\BibitemShut {NoStop}%
\bibitem [{\citenamefont {Rooney}\ \emph {et~al.}(2025)\citenamefont {Rooney}, \citenamefont {Luo}, \citenamefont {Stehouwer}, \citenamefont {Scappucci}, \citenamefont {Veldhorst},\ and\ \citenamefont {Jiang}}]{Rooney2025GateGermanium}%
  \BibitemOpen
  \bibfield  {author} {\bibinfo {author} {\bibfnamefont {J.}~\bibnamefont {Rooney}}, \bibinfo {author} {\bibfnamefont {Z.}~\bibnamefont {Luo}}, \bibinfo {author} {\bibfnamefont {L.~E.~A.}\ \bibnamefont {Stehouwer}}, \bibinfo {author} {\bibfnamefont {G.}~\bibnamefont {Scappucci}}, \bibinfo {author} {\bibfnamefont {M.}~\bibnamefont {Veldhorst}},\ and\ \bibinfo {author} {\bibfnamefont {H.-W.}\ \bibnamefont {Jiang}},\ }\bibfield  {title} {\bibinfo {title} {{Gate modulation of the hole singlet-triplet qubit frequency in germanium}},\ }\href {https://doi.org/10.1038/s41534-024-00953-3} {\bibfield  {journal} {\bibinfo  {journal} {npj Quantum Information}\ }\textbf {\bibinfo {volume} {11}},\ \bibinfo {pages} {1} (\bibinfo {year} {2025})}\BibitemShut {NoStop}%
\bibitem [{\citenamefont {Moutanabbir}\ \emph {et~al.}(2024)\citenamefont {Moutanabbir}, \citenamefont {Assali}, \citenamefont {Attiaoui}, \citenamefont {Daligou}, \citenamefont {Daoust}, \citenamefont {Vecchio}, \citenamefont {Koelling}, \citenamefont {Luo},\ and\ \citenamefont {Rotaru}}]{Moutanabbir2024NuclearWells}%
  \BibitemOpen
  \bibfield  {author} {\bibinfo {author} {\bibfnamefont {O.}~\bibnamefont {Moutanabbir}}, \bibinfo {author} {\bibfnamefont {S.}~\bibnamefont {Assali}}, \bibinfo {author} {\bibfnamefont {A.}~\bibnamefont {Attiaoui}}, \bibinfo {author} {\bibfnamefont {G.}~\bibnamefont {Daligou}}, \bibinfo {author} {\bibfnamefont {P.}~\bibnamefont {Daoust}}, \bibinfo {author} {\bibfnamefont {P.~D.}\ \bibnamefont {Vecchio}}, \bibinfo {author} {\bibfnamefont {S.}~\bibnamefont {Koelling}}, \bibinfo {author} {\bibfnamefont {L.}~\bibnamefont {Luo}},\ and\ \bibinfo {author} {\bibfnamefont {N.}~\bibnamefont {Rotaru}},\ }\bibfield  {title} {\bibinfo {title} {{Nuclear Spin-Depleted, Isotopically Enriched 70Ge/28Si70Ge Quantum Wells}},\ }\bibfield  {journal} {\bibinfo  {journal} {Advanced Materials}\ }\textbf {\bibinfo {volume} {36}},\ \href {https://doi.org/10.1002/adma.202305703} {10.1002/adma.202305703} (\bibinfo {year} {2024})\BibitemShut {NoStop}%
\bibitem [{\citenamefont {Chekhovich}\ \emph {et~al.}(2013)\citenamefont {Chekhovich}, \citenamefont {Makhonin}, \citenamefont {Tartakovskii}, \citenamefont {Yacoby}, \citenamefont {Bluhm}, \citenamefont {Nowack},\ and\ \citenamefont {Vandersypen}}]{Chekhovich2013NuclearDots}%
  \BibitemOpen
  \bibfield  {author} {\bibinfo {author} {\bibfnamefont {E.~A.}\ \bibnamefont {Chekhovich}}, \bibinfo {author} {\bibfnamefont {M.~N.}\ \bibnamefont {Makhonin}}, \bibinfo {author} {\bibfnamefont {A.~I.}\ \bibnamefont {Tartakovskii}}, \bibinfo {author} {\bibfnamefont {A.}~\bibnamefont {Yacoby}}, \bibinfo {author} {\bibfnamefont {H.}~\bibnamefont {Bluhm}}, \bibinfo {author} {\bibfnamefont {K.~C.}\ \bibnamefont {Nowack}},\ and\ \bibinfo {author} {\bibfnamefont {L.~M.}\ \bibnamefont {Vandersypen}},\ }\bibfield  {title} {\bibinfo {title} {{Nuclear spin effects in semiconductor quantum dots}},\ }\href {https://doi.org/10.1038/NMAT3652} {\bibfield  {journal} {\bibinfo  {journal} {Nature materials}\ }\textbf {\bibinfo {volume} {12}},\ \bibinfo {pages} {494} (\bibinfo {year} {2013})}\BibitemShut {NoStop}%
\bibitem [{\citenamefont {Xue}\ \emph {et~al.}(2019)\citenamefont {Xue}, \citenamefont {Watson}, \citenamefont {Helsen}, \citenamefont {Ward}, \citenamefont {Savage}, \citenamefont {Lagally}, \citenamefont {Coppersmith}, \citenamefont {Eriksson}, \citenamefont {Wehner},\ and\ \citenamefont {Vandersypen}}]{Xue2019BenchmarkingDevice}%
  \BibitemOpen
  \bibfield  {author} {\bibinfo {author} {\bibfnamefont {X.}~\bibnamefont {Xue}}, \bibinfo {author} {\bibfnamefont {T.~F.}\ \bibnamefont {Watson}}, \bibinfo {author} {\bibfnamefont {J.}~\bibnamefont {Helsen}}, \bibinfo {author} {\bibfnamefont {D.~R.}\ \bibnamefont {Ward}}, \bibinfo {author} {\bibfnamefont {D.~E.}\ \bibnamefont {Savage}}, \bibinfo {author} {\bibfnamefont {M.~G.}\ \bibnamefont {Lagally}}, \bibinfo {author} {\bibfnamefont {S.~N.}\ \bibnamefont {Coppersmith}}, \bibinfo {author} {\bibfnamefont {M.~A.}\ \bibnamefont {Eriksson}}, \bibinfo {author} {\bibfnamefont {S.}~\bibnamefont {Wehner}},\ and\ \bibinfo {author} {\bibfnamefont {L.~M.}\ \bibnamefont {Vandersypen}},\ }\bibfield  {title} {\bibinfo {title} {{Benchmarking Gate Fidelities in a Si/SiGe Two-Qubit Device}},\ }\href {https://doi.org/10.1103/PHYSREVX.9.021011} {\bibfield  {journal} {\bibinfo  {journal} {Physical Review X}\ }\textbf {\bibinfo {volume} {9}},\ \bibinfo {pages} {021011} (\bibinfo {year} {2019})}\BibitemShut {NoStop}%
\bibitem [{\citenamefont {Nielsen}\ \emph {et~al.}(2021)\citenamefont {Nielsen}, \citenamefont {Gamble}, \citenamefont {Rudinger}, \citenamefont {Scholten}, \citenamefont {Young},\ and\ \citenamefont {Blume-Kohout}}]{Nielsen2021GateTomography}%
  \BibitemOpen
  \bibfield  {author} {\bibinfo {author} {\bibfnamefont {E.}~\bibnamefont {Nielsen}}, \bibinfo {author} {\bibfnamefont {J.~K.}\ \bibnamefont {Gamble}}, \bibinfo {author} {\bibfnamefont {K.}~\bibnamefont {Rudinger}}, \bibinfo {author} {\bibfnamefont {T.}~\bibnamefont {Scholten}}, \bibinfo {author} {\bibfnamefont {K.}~\bibnamefont {Young}},\ and\ \bibinfo {author} {\bibfnamefont {R.}~\bibnamefont {Blume-Kohout}},\ }\bibfield  {title} {\bibinfo {title} {{Gate Set Tomography}},\ }\href {https://doi.org/10.22331/q-2021-10-05-557} {\bibfield  {journal} {\bibinfo  {journal} {Quantum}\ }\textbf {\bibinfo {volume} {5}},\ \bibinfo {pages} {557} (\bibinfo {year} {2021})}\BibitemShut {NoStop}%
\end{thebibliography}%

\end{document}